\documentclass[aps,prl,reprint,superscriptaddress,showpacs]{revtex4-1}
\usepackage{graphicx}
\usepackage{amssymb}
\usepackage{amsmath}
\usepackage{bm}
\usepackage{verbatim}
\usepackage[colorlinks=true,urlcolor=blue,citecolor=blue,linkcolor=blue]{hyperref}

\usepackage{enumitem} 

\begin{document}

\title{Phonon-mediated Casimir interaction between mobile impurities in one-dimensional quantum liquids}

\author{Michael Schecter}
\email[]{schecter@physics.umn.edu}
\affiliation{School of Physics and Astronomy, University of Minnesota,
Minneapolis, MN 55455, USA}

\author{Alex Kamenev}
\affiliation{School of Physics and Astronomy, University
of Minnesota, Minneapolis, MN 55455, USA}
\affiliation{William I. Fine
Theoretical Physics Institute, University of Minnesota, Minneapolis, MN
55455, USA}

\date{\today}

\begin{abstract}
Virtual phonons of a quantum liquid scatter off impurities and mediate a long-range interaction, analogous to the Casimir effect. In one dimension the effect is universal and the induced interaction decays as $1/r^3$, much slower than the van der Waals interaction $\sim1/r^6$, where $r$ is the impurity separation. The sign of the effect is characterized by the product of impurity-phonon scattering amplitudes, which take a universal form and have been seen to vanish for several integrable impurity models. Thus, if the impurity parameters can be independently tuned to lie on opposite sides of such integrable points, one can observe an attractive interaction turned into a repulsive one.
\end{abstract}


\maketitle

The concept of zero-point energy has fascinated minds ever since the inception of quantum mechanics. Beyond being merely an inconsequential redefinition of the ground state energy, it was shown by Casimir \cite{Casimir} that \emph{changes} in the zero-point energy can lead to observable forces between uncharged conducting plates. 
Going beyond simple planar geometries, recent developments in nanotechnology \cite{nano1,nano2,nano3,nano4,nano5,nano6,nano7,nano8,nano9} have stimulated intense efforts to understand Casimir interactions between conducting objects of arbitrary shape \cite{Kardar1,Kardar2}. 

Zero-point fluctuations of the electromagnetic field constitute only one example of a much broader class of phenomena. Essentially \emph{any} medium whose fluctuations display long-range correlations, \emph{e.g.,} media with a continuously broken symmetry and associated Goldstone mode(s) \cite{Kardar3}, induce long-range interactions between perturbing objects that modify the spectrum of fluctuations. A well-known example of such media is a superfluid whose Goldstone mode is the quantized sound mode or \emph{phonon}. Although $^4$He was the first system to exhibit the remarkable properties of superfluidity, recent advances in ultracold atom trapping and manipulation \cite{Bloch} have lead to an unprecedented ability to study ultra-clean bosonic or fermionic superfluids subject to tunable spatial dimensionality, lattice configuration and interaction strength. 

Perturbing objects or \emph{impurities} can be controllably introduced by transferring a fraction of atoms into a different hyperfine state \cite{Zwierlein,Kohl1,Schneble}, or by admixing a different atomic species \cite{Catani1,Kohl2,Denschlag,Catani2}. As we show, such impurities immersed in an {\em interacting}  cold atom environment is a particularly appealing setup. It gives rise to a long-range interaction, which we hereafter denote as the \emph{Casimir} interaction, due to scattering of virtual phonons, the same mechanism lying at the heart of the analogous photon-induced Casimir effect. In addition, the cold-atom analog of the interaction has the advantage of being continuously tunable both in magnitude and in {\em sign}, a task which is hardly achievable in a linear electromagnetic medium \cite{Pitaevskii_Casimir}.

In most studies of Casimir interactions the analysis is restricted to \emph{static} configurations of objects or bounding surfaces. The impurities in quantum liquids are typically free to propagate under the influence of the induced interactions. Therefore they must be regarded as \emph{mobile}, and in certain circumstances should be distinguished from their static counterparts characterized by infinite effective mass. As we discuss below, for a system of repulsively interacting fermions in 1d this leads to a \emph{qualitatively} different asymptotic behavior of the induced interaction between impurities.

Using an effective low-energy theory, we find a Casimir interaction between mobile impurities in a 1d quantum liquid given by
\begin{equation}
\label{eq:potential}
U_{\mathrm{Cas}}(r)=-mc^2\frac{\Gamma_1\Gamma_2}{32\pi}\frac{\xi^3}{r^3},
\end{equation}
where $m$ is the mass of particles in the fluid, $c$ is the sound velocity, $\xi=\hbar/mc$ and the dimensionless parameters $\Gamma_{1,2}$ are impurity-phonon scattering amplitudes discussed in detail below.

Owing to the enhanced role of fluctuations in 1d, the Casimir interaction exhibits a decay law $\sim1/r^3$  which is much slower than the van der Waals interaction between neutral atoms $U_{\mathrm{vdW}}(r)\sim1/r^6$ \cite{footnote4}. In accord with recent works on the single impurity problem \cite{Gang_Kam1,Gang_Kam2,Schec1,Matveev_Andreev}, we also find that $\Gamma$ acquires a universal form in terms of independently measurable thermodynamic characteristics. At special points in parameter space where the underlying model becomes integrable, the thermodynamics can be extracted exactly, thus yielding the scattering amplitude, which has been seen to vanish identically in several of these special cases, $\Gamma=0$ \cite{Gang_Kam1,Gang_Kam2,Schec1,Matveev_Andreev}. This has important implications for spinor condensates, which lie close to the integrable $SU(2)$ symmetric point, for which particles in different hyperfine states have equal masses and nearly equal interaction constants. It opens the possibility of observing an attractive interaction turned into a repulsive one when the impurity parameters are independently tuned to lie on opposite sides of the integrable point.

Our results should be contrasted with the works of  \cite{Recati1,Recati2,Schonhammer} who considered the limit of infinitely heavy impurities embedded in a 1d quantum liquid. They concluded that (a) repulsively interacting fermions mediate an interaction whose smooth component  \cite{footnote3} scales as $1/r$ and (b) repulsively interacting bosons (or attractively interacting fermions) do not mediate a smooth long-range interaction. In the case of {\em mobile} impurities we find that the induced interaction is different from (a) \& (b) and is instead given by Eq.~(\ref{eq:potential}). Moreover, even  \emph{static} impurities in  repulsively interacting bosonic liquids are subject to induced interactions of the form given by Eq.~(\ref{eq:potential}), in contradiction with (b).

Conclusions (a) \& (b) were reached by appealing to the scaling theory of the effective impurity strength \cite{Kane_Fisher}. The latter shows a non-trivial energy dependence in the presence of a quantum liquid due to its non-linear (\emph{i.e.,} interacting) nature. The physics behind this is that, in addition to direct scattering off the impurity, excitations of the liquid may also scatter indirectly off the local density distortion induced by the impurity. This mechanism gives rise to a  renormalization of the low-energy effective impurity strength: for a system of repulsively interacting fermions the impurity is renormalized into a perfectly reflecting barrier \cite{Kane_Fisher}, while for bosons the impurity instead becomes perfectly transparent (hence (b)).

However the Casimir effect comes {\em not} from the ultimate low-energy impurity strength, but rather relies on the participation of an \emph{energy band} of quantum fluctuations, whose width is set by $\hbar c/r$. As a result, there is a long-range Casimir force even in the case where the impurity strength renormalizes towards zero. Moreover such a flow towards zero impurity strength is a universal feature of mobile impurities in any environment, both bosonic and fermonic. This is because at energy scales below the recoil energy $E_{\mathrm{R}}\sim k_{\mathrm{F}}^2/M$ the impurity essentially 
decouples from (super-flows in) the host liquid \cite{Castro-Neto_Fisher}. This renders broad universality of our result -- Eq.~(\ref{eq:potential}) (the only exception being infinite mass objects in repulsive fermonic environment, interacting with
$1/r$ potential, \cite{Recati1,Recati2}).

We now develop the effective model leading to Eq.~(\ref{eq:potential}). In the absence of excitations, a system of impurities in a quantum liquid can be described by the Lagrangian $L_{\mathrm{imp}}=\sum _j [P_j\dot{X}_j-E_j(P_j,n)]$, where $X_j,\,P_j$ are impurity coordinates and momenta, respectively, and $E_j$ are the exact single-impurity dispersion relations, which incorporate renormalization due to the liquid \cite{Lamacraft}.

The relevant low-energy excitations of the quantum liquid are phonons \cite{Giamarchi} described by the density deviation $\rho$ from the background value $n$ and canonically conjugate superfluid phase $\varphi$, related to the superfluid velocity as $u=\partial_x\varphi/m$ ($\hbar=1$). Their dynamics are encoded in the phonon Lagrangian \cite{Popov,Haldane},
\begin{eqnarray}
\label{eq:phonon}
L_{\mathrm{ph}}=-\int_x\left[\rho\partial_t\varphi+\frac{n+\rho}{2m}(\partial_x\varphi)^2+\frac{mc^2}{2n}\rho^2+\frac{\alpha}{3!}\rho^3\right]
\end{eqnarray}
The quadratic (Luttinger liquid) part of the Lagrangian describes a linearly dispersing hydrodynamic mode, $\omega(q)=c|q|$. The cubic non-linear terms ~$\rho(\partial_x\phi)^2$ and ~$\rho^3$ are retained, as they turn out to be essential in deriving the correct impurity-phonon scattering amplitude $\Gamma$ \cite{Gang_Kam1,Gang_Kam2,Matveev_Andreev}.

The impurity-phonon coupling can then be derived by employing a "weak-coupling" expansion in phonon amplitude \cite{Castro-Neto_Fisher}. This description is valid at low energies where the impurity becomes essentially transparent. The expansion may be achieved by noting \cite{Baym_Ebner,Gang_Kam1,Gang_Kam2,Matveev_Andreev} that an impurity in the presence of a long wavelength phonon sees an essentially \emph{global} modification of the density $n+\rho(X)$ and supercurrent $u(X)$. One thus concludes that in the frame moving with velocity $u(X)$, the impurity energy is $E(P-Mu(X),n+\rho(X))$. A Galilean transformation then gives laboratory interaction energy: $E_\mathrm{lab}(P,u,n+\rho)=E(P-Mu,n+\rho)+Pu-\frac{1}{2}Mu^2$, where the phonon fields are taken at the location of the impurity. The left-hand side describes the additional energy cost associated with exciting the liquid in the presence of the impurity, thus expressing the impurity-phonon coupling through the exact impurity dispersion relation.

To describe scattering between impurities and phonons one may proceed along the lines of Refs.~\cite{Gang_Kam1,Gang_Kam2,Matveev_Andreev} (see the Supplemental Material \cite{supp} for details of this procedure): (i) perform a canonical transformation to variables $\tilde\rho,\,\tilde u$ which removes all coupling terms linear in phonon amplitude. (ii) expand the interaction energy $E_{\mathrm{lab}}(\tilde\rho,\tilde u)$ to second order in $\tilde\rho$, $\tilde u$. (iii) perform a rotation to the chiral basis defined by $\chi_\pm=\tilde{\vartheta}/\sqrt{\pi K}\pm\tilde{\varphi}\sqrt{K/\pi}$, where we introduced the displacement field $\tilde\vartheta$ through $\tilde\rho=\partial_x\tilde\vartheta/\pi$ and $K$ is the Luttinger parameter \cite{Giamarchi}. For repulsively interacting fermions (bosons) $K<1\,(K>1)$.

The first result of this sequence of operations is a spatially \emph{local} contribution to the inter-impurity potential mediated by single-phonon exchange \cite{footnote}. It is generated directly from the canonical transformation (i) and given by $U_{\mathrm{loc}}(X_i-X_j)=-c \gamma_{ij}\,\delta(X_i-X_j)$. Although the local interaction is not the main focus of this work, the explicit expression for $\gamma_{ij}$ is given in terms of partial derivatives of the dispersion in the Supplemental Material \cite{supp}.

More importantly, this procedure produces the fully renormalized scattering interaction between impurities and phonons, which ultimately leads to the indirect Casimir interaction between impurities. In the chiral phonon basis we find
\begin{eqnarray}
\label{eq:Lutt}L_{\mathrm{ph}}&=&\frac{1}{2}\sum_{\beta=\pm}\int_x\left[\chi_\beta(c\partial_x^2+\beta\partial_{x,t}^2)\chi_\beta\right];
\\
\label{eq:int}L_{\mathrm{int}}&=&\sum_j\frac{\Gamma_{j}}{m}\partial_x\chi_+(X_j,t)\partial_x\chi_-(X_j,t),\end{eqnarray}
where $\Gamma_j$ is the exact scattering amplitude corresponding to the $j^{\mathrm{th}}$ impurity. It represents the impurity-induced scattering matrix element between right and left moving chiral phonons \cite{footnote6}, and is discussed in more detail below.

\begin{figure}[b]
\includegraphics[width=0.93\columnwidth]{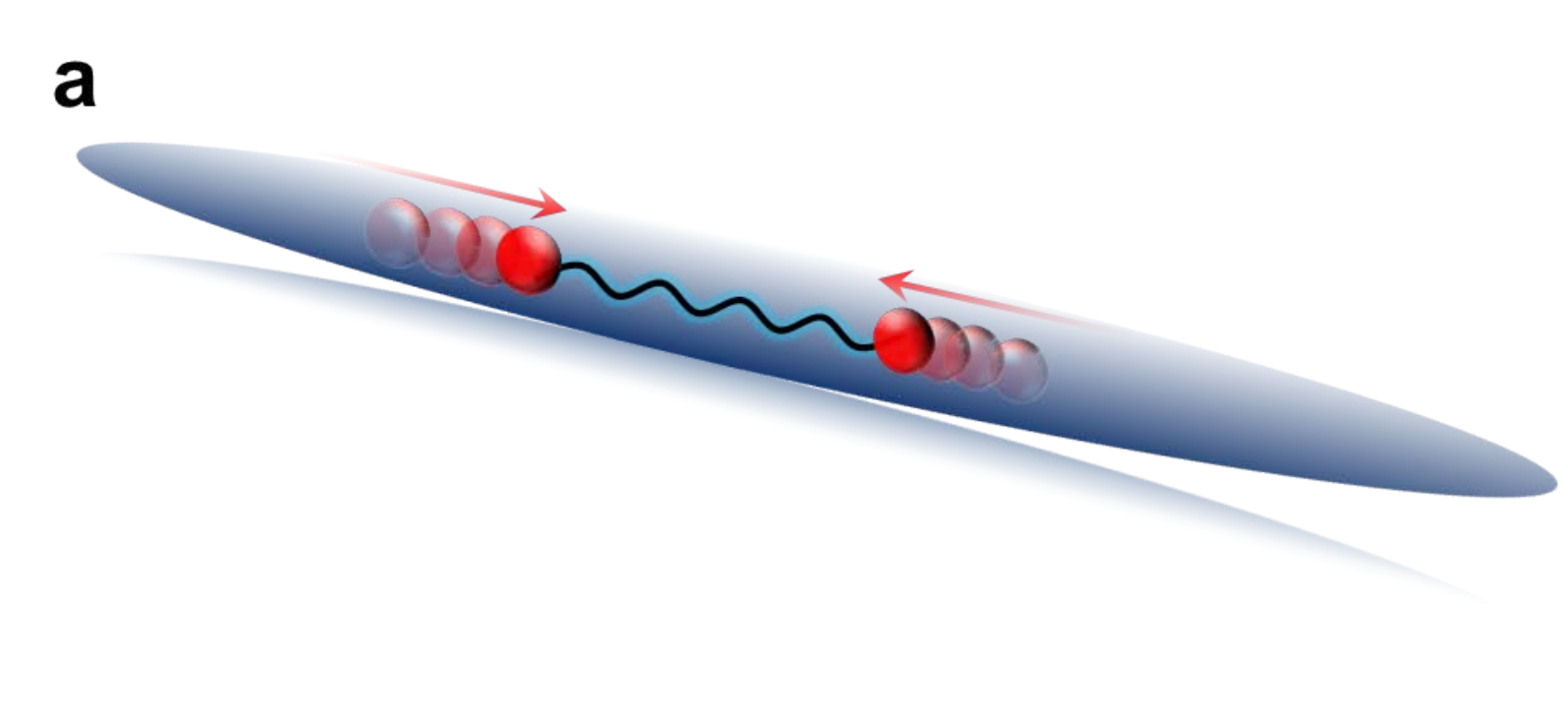}
\includegraphics[width=\columnwidth]{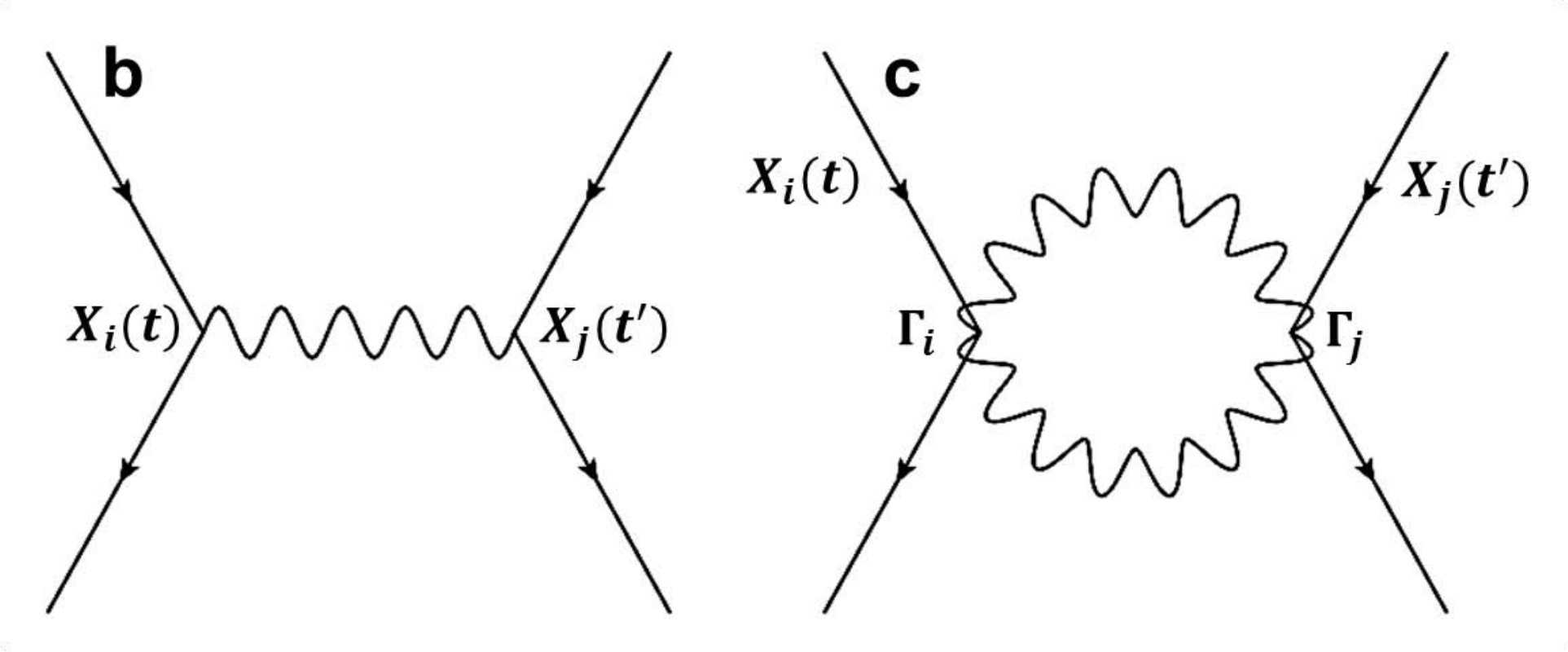}
\caption{(Color online) Casimir interaction between mobile impurities in a quantum liquid. \textbf{a}. Schematic depiction of two impurities experiencing an induced Casimir attraction mediated by phonon fluctuations of a one-dimensional quantum liquid. \textbf{b}. Single-phonon exchange (wavy line) leading to a spatially local inter-impurity interaction (impurities $i,\,j$ are denoted by straight lines with coordinates $X_{i,j}$). \textbf{c}. Two-phonon exchange responsible for the long-range Casimir interaction between impurities (see Eqs.~(\ref{eq:Lutt}), (\ref{eq:int}) for the corresponding Lagrangian).}
\label{fig:fig1}
\end{figure}

One may now evaluate the induced inter-impurity interaction by averaging the interaction, Eq.~(\ref{eq:int}), over the quadratic chiral action Eq.~(\ref{eq:Lutt}), see Fig.~(\ref{fig:fig1}) for a diagrammatic representation. This procedure gives the leading large distance, $r\gg\xi$, \cite{footnote7} component of the inter-impurity interaction, \emph{not} restricted to the regime of weak impurity coupling. It also incorporates the leading small temperature, $T\ll mc^2$, dissipative effects which are encoded in the semiclassical equations of motion,
\begin{eqnarray}
\label{eq:momentum}\dot{P}_i=-\sum_j \partial_{X_{i}}U\left(\frac{\dot{X}_i+\dot{X}_j}{2},X_{i}-X_j\right),
\\
\label{eq:U}U(V,X)=\frac{1}{2}\,\Gamma_{i}\Gamma_{j}\int\! \frac{dq}{2\pi}\,  \Pi(q,qV)\, \mathrm{e}^{iqX}.
\end{eqnarray}
Here $\Pi(q,\omega)$ is the polarization operator of the phonon gas, related to the Fourier transform of the retarded response function $\theta(t)\langle[\rho^2(x,t),\rho^2(0,0)]\rangle$, see \cite{supp}. The causality structure of $\Pi(q,\omega)$ allows one to express $U(V,X)$ in terms of even and odd components: $U=U_{+}+U_{-}$. Here $U_{\pm}(V,-X)=\pm U_{\pm}(V,X)$ correspondingly depend on the real and imaginary parts of $\Pi(q,\omega)$, which in turn are connected via the Kramers-Kronigs relation.

This decomposition is useful because it shows that the Casimir interaction explicitly stems from scattering of virtual excitations and is thus expressed through the real part of $\Pi$: $U_{\mathrm{Cas}}=U_+$. Evaluation of $U_+$ gives Eq.~(\ref{eq:potential}) at zero temperature. Below we discuss only the essential limiting behavior of $U_\pm$, delegating a section of the Supplemental Material \cite{supp} to the exact expressions.

The first effect of finite temperature is that the coherent nature of virtual phonon scattering is suppressed at separations beyond the temperature length $L_T=c/2\pi T$. This results in the exponential suppression of the Casimir interaction at large distances $r>L_T$,
\begin{equation}
\label{eq:potential_T}
U_{\mathrm{Cas}}(r)=-mc^2\frac{\Gamma_1\Gamma_2}{8\pi}\frac{\xi^3}{L_T^3}\mathrm{e}^{-2r/L_T}.
\end{equation}
The power-law scaling of Eq.~(\ref{eq:potential}), valid for $\xi<r<L_T$, is thus only meaningful deep in the quantum regime, $L_T\gg\xi$ or $T\ll mc^2$. 

The second effect is that \emph{real} excitations of the phonon background in general lead to impurity momentum relaxation, which manifests itself as damping in Eq.~(\ref{eq:momentum}). This is due to two-phonon Raman scattering and was studied extensively in Refs.~\cite{Landau_Khalatnikov,Baym_Ebner,Castro-Neto_Fisher,Gang_Kam1,Gang_Kam2,Schec1,Matveev_Andreev}. It may be seen in the simplest case of two slow, $\dot{X}_j\ll c$, symmetric impurities ($\Gamma_1=\Gamma_2=\Gamma$) where the equations of motion can be written in compact form using center of mass coordinates,

\begin{eqnarray}
\label{eq:eqs_motion}
\dot{p}=-\frac{\kappa}{2}\dot{r}-\frac{\partial U_{\mathrm{Cas}}(r)}{\partial r};\,\,\,\,\dot{P}=-2\kappa\dot{R}\left[1+f(r/L_T)\right].
\end{eqnarray}
Here $r=X_1-X_2$, $R=\frac{1}{2}(X_1+X_2)$, $p=\frac{1}{2}(P_1-P_2)$, $P=P_1+P_2$, $\dot{X}_i=\partial_{P_i}E(P_i)$ and $f(y)$ is a dimensionless function (related to $U_-$ \cite{supp}) with asymptotic behavior $f(y)=1$ for $y\ll1$ and $f(y)=-15y\mathrm{e}^{-2y}$ for $y\gg1$ \cite{Gang_Kam2}. The damping coefficient $\kappa$ is given by Eq.~(\ref{eq:kappa}) and is discussed below.

Equation~(\ref{eq:eqs_motion}) shows that in addition to standard terms of the form $\dot{P}_i=-\kappa\dot{X}_i$, there is also a correlation correction to the center of mass damping which emerges from coherent two-phonon exchange processes. A similar effect was derived in Ref.~\cite{Gang_Kam2} in the context of dark soliton dynamics and leads to an effective center of mass damping which depends on the relative separation: for $r\lesssim L_T$ the damping is essentially twice as strong, whereas for  $r\gtrsim L_T$ ($f<0$) it becomes slightly suppressed. Here we demonstrate that this a generic phenomena, not restricted to solitons in weakly interacting condensates.

The damping coefficient in Eq.~(\ref{eq:eqs_motion}) is given by 
\begin{equation}
\label{eq:kappa}
\kappa=(mc)^2\frac{2\pi^3\Gamma^2}{15}\left(\frac{T}{mc^2}\right)^4.
\end{equation}
The $T^4$ scaling of $\kappa$ (equivalent to the inverse linear response mobility) first appeared in the context of single impurity dynamics in Ref.~\cite{Castro-Neto_Fisher}. More recently it was discussed by Refs.~\cite{Gang_Kam1,Gang_Kam2,Schec1,Matveev_Andreev} where it was shown that for several cases the exact pre-factor is sensitive to deviations from integrability of the underlying model.

It is worth emphasizing that the exact coefficient entering Eq.~(\ref{eq:kappa}) involves the same parameter $\Gamma$ as in  Eqs.~(\ref{eq:potential}), (\ref{eq:potential_T}). This is because both the Casimir interaction and the damping coefficient are controlled by the same underlying scattering mechanism: the virtual excitations are responsible for $U_{\mathrm{Cas}}$ while the real processes lead to collective damping of the center of mass. Remarkably, the two effects are related by the causality structure inherent in Eq.~(\ref{eq:U}).

The other main achievement of the theory is that it allows one to express the exact scattering amplitude $\Gamma$ in terms of partial derivatives of the exact single-impurity dispersion relation \cite{Gang_Kam1,Gang_Kam2,Schec1,Matveev_Andreev} (this follows from step (ii) above). The general expression for $\Gamma$ is not essential to the present discussion and can be found in previous works \cite{Gang_Kam1,Gang_Kam2,Schec1,Matveev_Andreev} as well as in the Supplemental Material \cite{supp}. What is crucial is the observation that for several physically relevant models $\Gamma$ changes sign across integrable points in parameter space. 

To demonstrate this we briefly discuss the weakly interacting Bose gas with impurities having nearly the same mass and coupling constants $G$ as the background gas $g$ \cite{footnote1}, relevant to spinor Bose condensates. In that case the system is near the integrable $SU(2)$ symmetric point ($M=m,\,G=g$) known as the Yang-Gaudin model \cite{Yang_Yang,Gaudin}. In this case it has been shown that $\Gamma=\frac{G}{c}\left(mG/Mg-1\right)$ (see the Supplemental Material \cite{supp} or Refs.~\cite{Schec1,Gang_Kam1}). One thus sees that if two separated impurities can be independently tuned to lie on opposite sides of integrability (say with $M=m$, $G_1>g$ and $G_2<g$), one can achieve a \emph{repulsive} Casimir interaction, instead of an attractive one.

The 
analysis utilized above also follows through entirely for \emph{static} impurities in 1d interacting \emph{bosonic}  (or attractively interacting fermionic) systems. 
This offers possibly the most straightforward way to verify the $1/r^3$ law (although in this case, to authors' knowledge, one does not have a possibility of tuning through integrability, as in the case of mobile impurities). To this end one would  pin two impurites with the help of a state-dependent optical lattice \cite{Schneble} or a species selective dipole potential \cite{Catani2}, and perform rf spectroscopy on individual impurity atoms \cite{Recati1,Esslinger}. As a function of their separation, one can then measure the corresponding line shifts of suitable internal hyperfine energy levels.

To estimate the magnitude of the Casimir effect we focus on the strongly-interacting Tonks limit where the energy scale $mc^2$ is largest. For the experiment of Ref.~\cite{Catani2} one finds a density of $^{87}$Rb atoms $n\approx 7\,(\mu\mathrm{m})^{-1}$ with a typical interaction strength of $mg/\hbar^2 n\approx 1$. With temperatures as low as $T=300\,\mathrm{nK}$ and a speed of sound $c\approx 1\,\mathrm{cm/s}$ this leads to $mc^2/\hbar\approx 138\,\mathrm{kHz}$ and $k_{\mathrm{B}}T/mc^2\approx 0.29$. Thus the magnitude of the potential at the closest applicable separation $r=\xi\approx1/n\approx 0.14\,\mu\mathrm{m}$ is $U_{\mathrm{Cas}}(\xi)\approx -1\,\mathrm{kHz}$, while for $r=5 L_T\approx 0.39\,\mu\mathrm{m}$ one finds $U_{\mathrm{Cas}}(5L_T)\approx -1\,\mathrm{Hz}$, indicating 3 orders of magnitude variation over a $\sim0.25\,\mu\mathrm{m}$ range of separations. The magnitude of the effect $\sim 1\,\mathrm{kHz}$ is thus within an experimentally accessible range, with the scale of applicable separations increasing as one goes deeper into the quantum degenerate regime, $T\ll mc^2$.

In conclusion, we have shown that  mobile impurities immersed in a 1d quantum liquid are subject to a long-range Casimir interaction, universally given by Eq.~(\ref{eq:potential}). This happens in spite of the fact that they become transparent at low energies. For several integrable impurity models, the amplitude of impurity-phonon scattering vanishes, leading to the absence of Casimir interactions in those systems. Finally, the strength of the effect is estimated to be within the resolution of current cold atom experiments, opening the possibility of observing the Casimir effect in a highly tunable, non-linear environment.

We thank D. Gangardt, D. Schneble and A. Abanov for useful discussions. M. S. acknowledges support from the University of Minnesota as well as the warm hospitality of Professor D. Schneble's group at Stony Brook University. A. K. and M. S. were supported by DOE Contract No. DE-FG02-08ER46482. 


\onecolumngrid
\section{{\Large Supplemental Material}}

In this Supplemental Material we present the calculation corresponding to steps (i)\--(iii) in the main text and include explicit expressions for the parameter $\gamma_{ij}$ as well as the scattering amplitude $\Gamma$. Detailed expressions (including various limits) for the function $U(V,X)=U_+(V,X)+U_-(V,X)$ entering Eq.~(\ref{eq:U}) of the main text are also presented.

\section{Calculation leading to Eq.~(\ref{eq:int}) of the main text}

We start from Lagrangian presented in the main text: $L=L_{\mathrm{imp-ph}}+L_{\mathrm{ph}}$ where

\begin{eqnarray}
\label{eq:lag1}
L_{\mathrm{imp-ph}}&=&\sum_j\left[P_j(\dot{X}_j-u_j)-E_j(P_j-M_ju_j,n+\rho_j)\right.
+\left.\frac{1}{2}M_ju_j^2\right],
\\
\label{eq:lag2}
L_{\mathrm{ph}}&=&-\int_x\left[\rho\partial_t\varphi+\frac{n+\rho}{2m}(\partial_x\varphi)^2+\frac{mc^2}{2n}\rho^2+\frac{\alpha}{3!}\rho^3\right],
\end{eqnarray}
where $\rho_j,u_j$ are phonon fields taken at the location of the $j^{\mathrm{th}}$ impurity. We now perform a canonical transformation to remove terms linear in phonon amplitude,
\begin{eqnarray}
\label{eq:1}
\tilde\rho&=& \rho+\sum_j N_j\delta(x-X_j),
\\
\label{eq:2}
\tilde{u}&=& u+ \frac{1}{m}\sum_j \Phi_j\delta(x-X_j),
\\
\label{eq:3}
\tilde{P}_j&=& P_j-\Phi_j \tilde\rho_j - m N_j\tilde{u}_j.
\end{eqnarray}
Here $N_j,\,\Phi_j$ are yet undetermined parameters which, as we discuss below, correspond respectively to the number of depleted particles and phase drop induced by the $j^{\mathrm{th}}$ mobile impurity \cite{Schec1}. Substituting Eqs.~(\ref{eq:1}-\ref{eq:3}) into Eqs.~(\ref{eq:lag1}), (\ref{eq:lag2}) and expanding to the second order in $\tilde\rho,\tilde{u}$ gives (suppressing tildes) $L=L_{\mathrm{imp}}+L_{\mathrm{ph}}+\sum_j L_{\mathrm{int},j}-\frac{1}{2}\sum_{i\neq j} U_{\mathrm{loc}}$, where $U_{\mathrm{loc}}$ is given below and
\begin{eqnarray}
\label{eq:int1}
\notag L_{\mathrm{int},j}&=&-\left(P_j-n\Phi_j-(M_j-mN_j)\dot{X}_j\right)u_j -\left(\partial_n E_j-(mc^2/n) N_j+\Phi_j \dot{X}_j\right)\rho_j
\\
&-&\frac{1}{2}(\rho_j,u_j)\hat{E}_j^{\prime\prime}\left(\begin{array}{c}
\rho_j\\
u_j\end{array}\right).
\end{eqnarray}
One now sees that the terms linear in $\rho,\,u$ can be eliminated by demanding that $N_j,\,\Phi_j$ satisfy the equations: $P_j=n\Phi_j+(M_j-mN_j)\dot{X}_j,\,\partial_n E_j=(mc^2/n)N_j-\Phi_j\dot{X}_j$, which appeared in Ref.~\cite{Schec1}.

The second order terms are written in terms of the Hessian matrix (suppressing index $j$),
\begin{eqnarray}
\label{eq:hessian}
\hat{E}^{\prime\prime}=
\left[\begin{array}{cc}
\partial_{n}^{2}(E-\mu N)+2\Phi\partial_{n}V+\frac{\Phi^{2}}{M^{*}}, & -\frac{M-mN}{M^{*}}(\Phi+2M^{*}\partial_{n}V)\\
-\frac{M-mN}{M^{*}}(\Phi+2M^{*}\partial_{n}V), & \frac{(M-mN)(M-mN-M^{*})}{M^{*}}\end{array}\right],
\end{eqnarray}
where $V(P,n)=\partial_P E(P,n)$, $M^*=[\partial^2_P E(P,n)]^{-1}$ and $\mu(n)$ is the chemical potential of background particles. Finally, after rotation to the chiral phonon basis, $\chi_\pm=\vartheta/\sqrt{\pi K}\pm\varphi \sqrt{K/\pi}$, Eq.~(\ref{eq:int1}) reduces to Eq.~(\ref{eq:int}) of the main text. The off-diagonal component of the rotated Hessian matrix is thus identified with the scattering amplitude $\Gamma$, given below.

\subsection{Expression for $\gamma_{ij}$ of the local interaction}

In addition to the two-phonon scattering processes which lead to the long-range components described below by Eqs.~(\ref{eq:U-}$\--$\ref{eq:U-2}), we also find a spatially local contribution due to single phonon exchange. This is generated by the canonical transformation above and leads to a spatially local interaction potential of the form $U_{\mathrm{loc}}(X_i-X_j)=-c\gamma_{ij}\delta(X_i-X_j)$ where
\begin{eqnarray}
\label{eq:loc}
\gamma_{ij}&=&\frac{\delta_{i,+}\delta_{j,+}}{2\pi}\left(1-\frac{V_i+V_j}{2c}\right)+\frac{\delta_{i,-}\delta_{j,-}}{2\pi}\left(1+\frac{V_i+V_j}{2c}\right).
\end{eqnarray}
Here $\frac{\delta_\pm}{\sqrt\pi}=-\sqrt{\frac{K}{\pi}}\Phi \mp \sqrt{\frac{\pi}{K}}N$ are the phonon scattering phase shifts induced by the mobile impurity. As shown previously in Refs.~\cite{Zvon_Chei_Giamarchi2,Imambekov_Glaz1,Imambekov_Glaz2,Imambekov_Glaz3,Kam_Glaz}, the scattering phase shifts may be expressed in terms of the thermodynamic characteristics, \emph{i.e.,} partial derivatives of the exact impurity dispersion relation $E(P,n)$. By solving for $N$ and $\Phi$ above (with $\dot{X}=V$) we find
\begin{eqnarray}
\label{eq:delta}\frac{\delta_\pm}{\pi}=\mp\frac{1}{\sqrt{K}}\frac{1}{1\mp V/c}\left[\frac{n\partial_n E}{mc^2}\pm\frac{P-MV}{mc}\right],
\end{eqnarray}
which acquires the form presented in Refs.~\cite{Imambekov_Glaz1,Imambekov_Glaz2,Imambekov_Glaz3,Kam_Glaz}.

\subsection{Expression for $\Gamma$}

As first shown in Refs.~\cite{Gang_Kam1,Gang_Kam2,Schec1,Matveev_Andreev}, the scattering amplitude $\Gamma$ can be expressed in terms of partial derivatives of the exact dispersion. Since $\delta_\pm$ also depend on partial derivatives of the same dispersion, one can partially simplify the expression for $\Gamma$  by using together Eqs.~(\ref{eq:hessian}) \& (\ref{eq:delta}). This yields a relatively compact form for the scattering amplitude,
\begin{eqnarray}
\label{eq:Gamma}
\notag\Gamma&=&-\frac{n}{c}\partial^2_n\left(E-\frac{M}{m}\mu\right)-\frac{m}{M^*}\frac{\tilde\delta_+\tilde\delta_-}{\pi}
-m\sqrt{K}\left[\frac{\tilde\delta_+}{\pi}\partial_n(c-V)-\frac{\tilde\delta_-}{\pi}\partial_n(c+V)\right],
\end{eqnarray}
where $\tilde\delta_\pm=\delta_\pm\pm\frac{M}{m}\frac{\pi}{\sqrt{K}}$. At small momentum one finds
\begin{eqnarray}
\label{eq:Gamma1}
\Gamma=\frac{n}{c}\partial_n^2 \tilde\mu-\frac{n}{M^* c^3}\left(\partial_n\tilde\mu\right)^2-\frac{2n}{c^2}(\partial_n c)(\partial_n \tilde\mu),
\end{eqnarray}
where $\tilde\mu=E(0,n)-\frac{M}{m}\mu$. For a generic $SU(2)$ symmetric gas with equal coupling constants and masses ($M=m$), one has $E(0,n)=\mu(n)\implies\tilde\mu=0$, and thus $\Gamma=0$ (note that in this case $\Gamma=0$ actually holds for \emph{arbitrary} momenta, as can be checked using the exact Bethe Ansatz solution for $E(P,n)$). For a gas of weakly interacting particles with local interactions of strength $g$ between background particles and $G$ between background and impurity, one finds $\mu=ng=mc^2$, $E(0,n)=nG$ and $M^*\approx M$ to leading order in $g,\,G$. Using Eq.~(\ref{eq:Gamma1}) then gives $\Gamma=\frac{G}{c}\left(mG/Mg-1\right)$, as stated in the main text. One thus sees explicitly that $\Gamma\to0$ at the $SU(2)$ symmetric point: $M=m$ and $G=g$. For the functional form of $\Gamma$ for other coupling strengths and models, we refer to the expressions derived in Refs.~\cite{Gang_Kam1,Gang_Kam2,Schec1,Matveev_Andreev}.

\section{Expression for $U_{\pm}$}

\begin{figure}[t!]
\includegraphics[width=.495\columnwidth]{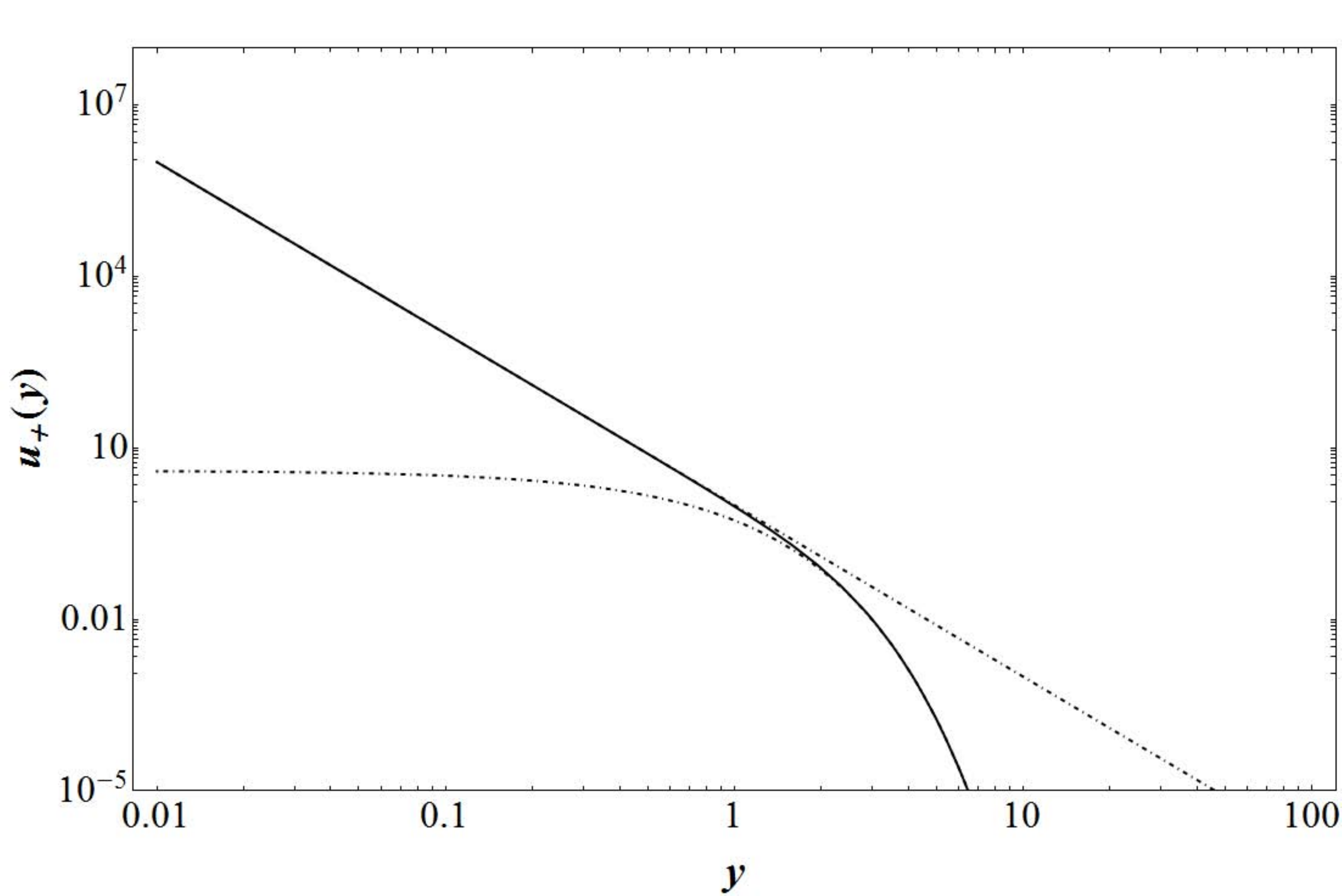}
\includegraphics[width=.495\columnwidth]{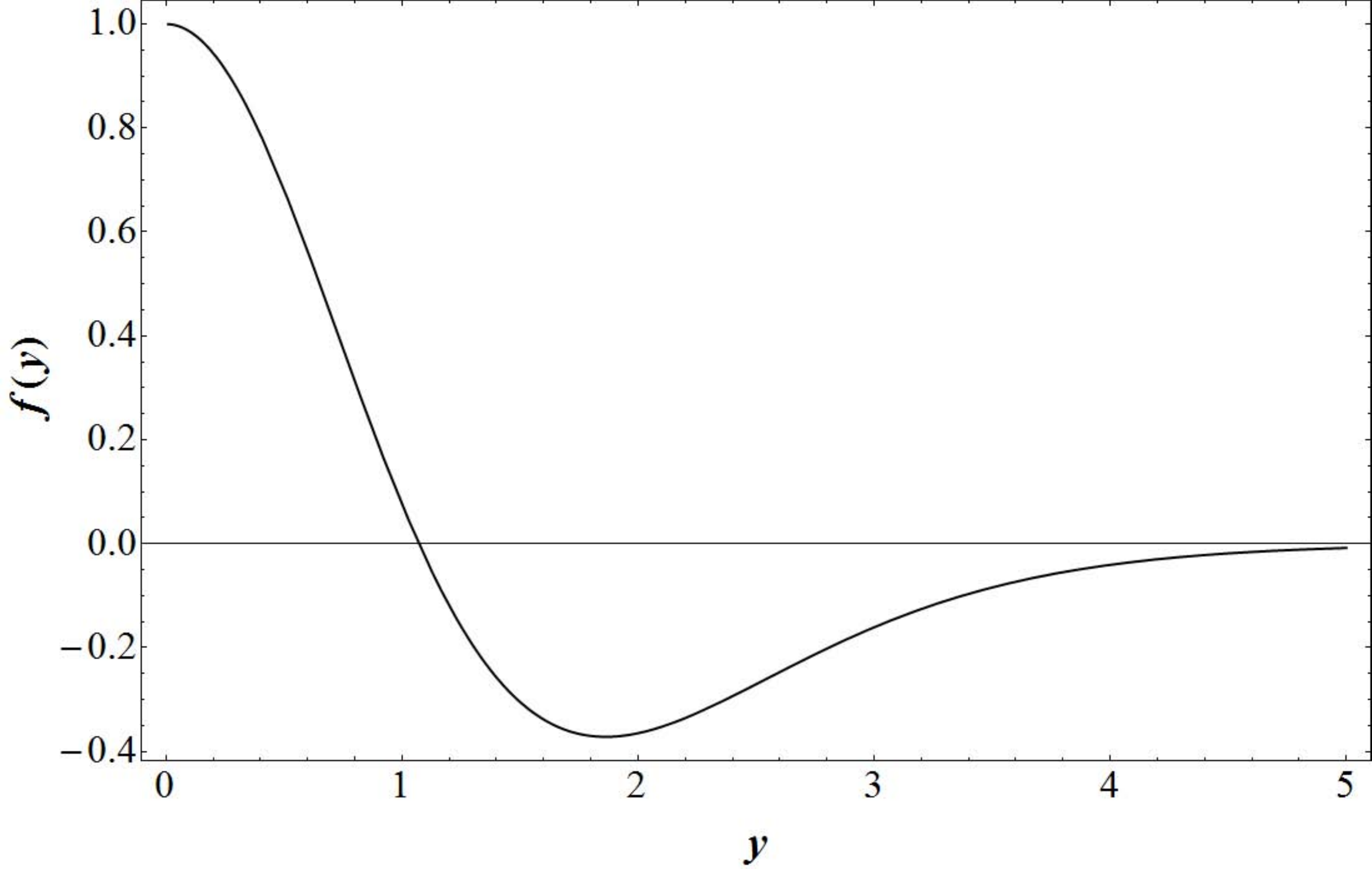}
\caption{Top panel: Log-log plot of the function $u_+(y)$ (solid curve, Eq.~(\ref{eq:U+2})). It interpolates between the asymptotic limits: $u_+(y)=1/y^3$ for $y\ll1$ and $u_+(y)=4\mathrm{e}^{-2y}$ for $y\gg1$ (dashed line and curve, respectively) described by Eqs.~(\ref{eq:potential}) and (\ref{eq:potential_T}) of the main text. Bottom panel: Dimensionless function $f(y)$, Eq.~(\ref{eq:f}), controlling the correlation correction to the center of mass friction, see Eqs.~(\ref{eq:eqs_motion}) of the main text.}
\label{fig:fig2}
\end{figure}

As briefly mentioned in the main text, the Casimir interaction potential is related to $U_+(V,X)$ while the friction is related to $U_-(V,X)$. This decomposition is particularly convenient because $U_{\pm}$ only depend, respectively, on the real and imaginary parts of the 
retarded phonon polarization operator $\Pi(q,\omega)$, which possesses the causality structure. As a result, the real part can be extracted from the imaginary part (or vice versa) using the Kramers-Kronigs relation $\mathrm{Re}\Pi(q,\omega)=\frac{1}{\pi}\int\frac{d\omega^\prime}{\omega^\prime-\omega}\mathrm{Im}\Pi(q,\omega^\prime)$. The imaginary part is given by
\begin{equation}
\label{eq:pol}
\mathrm{Im}\Pi(q,\omega)=\frac{q^2-\omega^2/c^2}{64c}\left[\mathrm{coth}\frac{cq+\omega}{4T}-\mathrm{coth}\frac{cq-\omega}{4T}\right].
\end{equation}
Restricted to the impurity trajectory, $\omega=qV$, one sees that $\mathrm{Im}\Pi(q,qV)$ is an odd function of $q$ while the real part is an even function. From Eq.~(\ref{eq:U}) of the main text, we then have $U_+(V,X)\propto\int_q\mathrm{cos}(qX)\mathrm{Re}\Pi(q,qV)$ and $U_-(V,X)\propto\int_q\mathrm{sin}(qX)\mathrm{Im}\Pi(q,qV)$. Using Eq.~(\ref{eq:pol}) we obtain
\begin{eqnarray}
\label{eq:U+}
&U&_+(V,X)=-mc^2\frac{\Gamma_i\Gamma_j}{64\pi}\left(1-V^2/c^2\right)\left[\frac{1}{1-V/c}\frac{\xi^3}{L_+^3}\frac{\mathrm{cosh}\frac{X}{L_+}}{\mathrm{sinh}^3\frac{|X|}{L_+}}+\frac{1}{1+V/c}\frac{\xi^3}{L_-^3}\frac{\mathrm{cosh}\frac{X}{L_-}}{\mathrm{sinh}^3\frac{|X|}{L_-}}\right],
\end{eqnarray}
\begin{eqnarray}
\label{eq:U-}U_-(V,X)=-mc^2\frac{\Gamma_i\Gamma_j}{64\pi}\left(1-V^2/c^2\right)\left[\frac{\xi^3}{L_-^3}\frac{\mathrm{cosh}\frac{X}{L_-}}{\mathrm{sinh}^3\frac{X}{L_-}}-\frac{\xi^3}{L_+^3}\frac{\mathrm{cosh}\frac{X}{L_+}}{\mathrm{sinh}^3\frac{X}{L_+}}\right].
\end{eqnarray}
Here $L_\pm=\frac{c\mp V}{2\pi T}$ are the Doppler shifted thermal lengths in the co-moving (+) and counter-propagating ($-$) frames. The leading order expansion in small velocity $V/c\ll 1$ is given by $U_+(0,X)=-mc^2\frac{\Gamma_i\Gamma_j}{32\pi}\frac{\xi^3}{L_T^3}u_+(X/L_T)$ and $U_-(V,X)=-mc^2\frac{\Gamma_i\Gamma_j}{120\pi}\frac{V}{c}\frac{\xi^3}{L_T^3}u_-(X/L_T)$, where
\begin{eqnarray}
\label{eq:U+2}
u_+(y)&=&\frac{\mathrm{cosh}y}{\mathrm{sinh}^3 y},
\\
\label{eq:U-2} u_-(y)&=&\frac{45}{4}\frac{y\left(\mathrm{coth}^2 y-\frac{1}{3}\right)-\mathrm{coth}y}{\mathrm{sinh}^2 y}.
\end{eqnarray}
The inter-impurity potential is given by $U_{\mathrm{Cas}}=U_+(0,r)$, which is related to Eq.~(\ref{eq:U+2}) and interpolates between the asymptotic limits described by Eqs.~(\ref{eq:potential}) and (\ref{eq:potential_T}) of the main text, see Fig.~(\ref{fig:fig2}). The function $f(y)$ entering Eq.~(\ref{eq:eqs_motion}) of the main text is given by $f(y)=\partial_y u_-(y)$, see Fig.~(\ref{fig:fig2}). Using Eq.~(\ref{eq:U-2}), one finds
\begin{equation}
\label{eq:f}
f(y)=15\,\mathrm{csch}^2 y\left[1+\left(1-y\,\mathrm{coth}y\right)\left(1+3\,\mathrm{csch}^2 y\right)\right],
\end{equation}
so that $f$ changes sign at $y=r/L_T=1.07$.



\begin{thebibliography}{26}
\bibitem{footnote4} Based on a phase-space argument we expect that in dimension $d$ the induced interaction scales as $U_{\mathrm{Cas}}(r)\sim1/r^{2d+1}$, see Eq.~(\ref{eq:U}). In this work we focus on the case $d=1$ where, in addition to the Casimir effect being strongest at large separations, integrability of the underlying model also plays an important role.
\bibitem{footnote7} Our low-energy description is valid when the bandwidth, $\hbar c/r$, and temperature are smaller than $mc^2$ and $E_{\mathrm{R}}$ (the latter inequality applying only for $K<1$). For systems near integrability one has $M\sim m$ so that $E_{\mathrm{R}}\sim E_{\mathrm{F}}\gtrsim mc^2$. Hence Eq.~(\ref{eq:potential}) is applies only for $r>\xi$. Discussion of the full cross-over behavior for $M\gg m$ and $K<1$ is beyond the scope of this work.
\end{thebibliography}
\end{document}